# Maximizing Infrared Transmission Contrast Upon Phase Transition of Thermally Grown Vanadium Dioxide Thin Films by Rapid Thermal Processing

*Ken Araki, Vishwa Krishna Rajan, and Liping Wang\**

K. Araki, V. K. Rajan, L. Wang

School for Engineering of Matter, Transport and Energy, Arizona State University, 501 E Tyler Mall, Tempe, AZ, 85287, USA

E-mail: liping.wang@asu.edu

Funding: NSF Grant No. CBET-2212342

**Abstract**: Pristine vanadium dioxide ($VO_2$), an insulator-to-metal transition (IMT) material, is grown via furnace oxidation followed by rapid thermal annealing with forming gas (5%$H_2$ / 95%$N_2$) which reduces surface over-oxides such as $V_2O_5$ formed during the oxidation. The evolutional IMT behaviors of the thermochromic film and vanadium oxide states over different reduction time are systematically studied with temperature-dependent infrared spectrometry, electrical resistivity, and X-ray diffraction measurements. After optimally reducing surface over-oxides to $VO_2$, infrared transmission contrast upon phase transition is enhanced to 46% (at 9 μm wavelength) compared to 23% from fully oxidation without any reduction. Moreover, pristine $VO_2$ thin film obtained from thermal oxidation and optimal reduction processes exhibits sharp phase transition and narrow thermal hysteresis within 2~4°C in both infrared transmission and electrical resistivity, which are comparable to the $VO_2$ of best quality prepared by other sophisticated fabrication techniques. The thermally grown method presented here would facilitate the scalable fabrication of high-quality $VO_2$ thin films and tunable radiative coatings for high-performance thermal control applications.

1. Introduction

Tunable optical and infrared properties of radiative coatings are highly desired for self-adaptive solar thermal energy harvesting and radiative thermal control to the changing environment. Vanadium dioxide ($VO_2$) as one of the most popular thermochromic materials has attracted lots of attention in such applications due to its unique insulator-to-metal transition (IMT).[1] Tunable $VO_2$ coatings using Fabry-Perot cavity can switch from high reflective insulating phase to highly emissive metallic phase with an increase in temperature for passive radiative thermal control.[2-5] Other $VO_2$ coatings can be employed as a radiative cooler during night time while it can behave as a selective solar absorber at day time.[6-8] Tunable bandpass filters, which have the ability to pass both narrowband and broadband mid-infrared light only at insulating phase, are also designed based on $VO_2$ phase transition.[9, 10] Due to its transparency at insulating phase, studies have also demonstrated anti-reflection in mid-infrared region by depositing thin layer of $VO_2$ on epsilon near-zero substrates.[11, 12] Such technique can be used as a thermal rectifier that allow large difference between forward and reverse heat flow upon the phase transition temperature using $VO_2$ on doped silicon, demonstrated in far-field regime[13] and on sapphire demonstrated in near-field regime.[14]

Established methods to fabricate $VO_2$ thin films include molecular beam epitaxy (MBE),[15, 16] reactive sputtering,[17-20] atomic layer deposition (ALD),[21-23] pulsed laser deposition (PLD),[24, 25] sol-gel,[26] hydrothermal process,[27, 28] and thermal oxidation[29, 30]. However, some growth techniques that provide best quality of $VO_2$ suffer from sophisticated equipment, expensive operation, slow deposition rate, or highly selective lattice-matched substrates, which greatly hinders the scalability. Other cost-effective techniques in particular thermal oxidation bears compromised film quality due to surface over-oxidation with vanadium pentoxide ($V_2O_5$) in oxygen-rich atmosphere, which causes unwanted phonon absorption, curtailing the contrast in the infrared properties upon phase transition.[31] It has been demonstrated that with longer oxidation time, XRD intensity of $VO_2$ decreases while $V_2O_5$ signal increases as the vanadium film is oxidized in the air at atmospheric pressure.[32] With rapid thermal oxidation, formation of $V_2O_5$ can be minimized but it requires precise time control from several seconds to one minute.[33] Oxidation at high temperatures without control in oxygen partial pressure could result in high ratio of $V_2O_5$.[34] However, with precisely-controlled oxygen and hydrogen concentrations, pristine $VO_2$ can be obtained as a result of a balance between over-oxidation and reduction.[35] High quality $VO_2$ can be synthesized directly from further oxidizing $V_2O_3$ in well-controlled oxygen partial pressure.[36] On the other hand, $VO_2$ can also be obtained from reducing ALD prepared high-oxides ($V_xO_y$ with y/x >2) or $V_2O_5$ annealed at high temperatures for long hours in extremely low-oxygen environment,[37, 38] which improves $VO_2$ solar transmission contrast for smart windows[39] and its electrical resistivity contrast.[40]

To facilitate the reduction of surface over-oxides, hydrogen gas can be carefully introduced either directly or in the forming gas (5%$H_2$ / 95%$N_2$) to reactively remove redundant oxygen atoms in the film at high temperatures.[41] The use of forming gas can result in denser crystallization at higher temperatures and at longer hours.[42] Note that similar reduction via solution-based chemical reaction using hydrogen-based agents has been employed to study monoclinic $VO_2$ (M) from $V_2O_5$ nanopowders and nanoparticles, known as hydrothermal process.[27] However, as hydrothermal synthesis requires optimization of multiple chemical reaction processes during which $VO_2$ phase undergoes transformation into metastable phase such as $VO_2$ (B) and $VO_2$ (A), it is challenging to obtain pure monoclinic phase $VO_2$ (M)[43, 44]

In this work, we demonstrate maximal infrared transmission contrast from pristine $VO_2$ thin film prepared by cost-effective thermal growth method involving sputtering, furnace oxidation in $O_2$-rich atmosphere, and surface reduction with forming gas via rapid thermal processing. Vanadium thin films of different thicknesses are sputtered onto undoped silicon wafers. Oxidation time study at 350°C is carried out to determine the fully oxidation condition with minimal surface over-oxides. Reduction time study at 500°C is performed to reach the optimal reduction where pristine $VO_2$ is obtained. Temperature-dependent infrared transmittance and electrical resistivity are characterized to show the evolutional IMT behaviors of the oxidized vanadium thin film with different oxide states over reduction time, which are directly revealed by X-ray diffraction measurements. Finally, complex refractive indices of pristine $VO_2$ thin film from this thermal growth method are fitted for both insulating and metallic phases with the help of global optimization.

## 2. Results and Discussion

Fabrication process to obtain high quality thermally grown VO$_2$ thin films consists of vanadium deposition via sputtering, thermal oxidation in O$_2$-rich furnace, and surface reduction via rapid thermal annealing with forming gas, as depicted in Fig. 1 (See Sec. 4 for experimental details). To begin with, vanadium films of different thicknesses are deposited with DC sputtering (Fig. 1a) onto double-side polished undoped silicon wafers. Three different thicknesses of vanadium are deposited separately (100 nm, 150 nm, and 200 nm) which are studied for optimal oxidation time to determine largest possible infrared transmittance contrast. Next, the sputtered vanadium thin films are thermally oxidized at 350°C in a muffle furnace rich of O$_2$ (Fig. 1b). Insufficient oxidation will leave some vanadium unoxidized, which leads to lower transmission in the insulating phase. Excessive oxidation will lead to high-oxides such as V$_2$O$_5$ on the film surface, which would yield higher transmission in the metallic phase. In this study, optimal oxidation time is determined when the infrared transmittance at insulating phase saturates. Note that under this fully oxidation condition, all the vanadium should be just oxidized, while there still exists some amount of surface over-oxides due to large O$_2$ concentration in the furnace atmosphere. Lastly, the fully oxidized vanadium thin films are taken for surface reduction at 500°C in rapid thermal processing with 0.1 LPM forming gas (5%H$_2$ / 95%N$_2$) as a reduction agent (Fig. 1c). The reduction time is further studied to achieve the best quality of VO$_2$ thin films with all surface over-oxides reduced to VO$_2$ without over-reducing it to lower-oxides or vanadium, under which the infrared transmission contrast is maximized upon VO$_2$ phase transition. Temperature-dependent infrared spectrometry, temperature-dependent electrical resistivity measurements, and X-Ray diffraction characterizations (See Sec. 4 for experimental details) are carried out to fully understand the effects of oxidation time and reduction time on the oxidation states, IMT phase transition behaviors, and film compositions. In addition, the film thickness is also measured with profilometry for 150-nm-thick vanadium thin film after sputtering (150±2.6 nm), fully oxidation (315±14 nm), and optimal reduction (304±16 nm) (See Fig. S1) for fitting the complex refractive indices of thermally grown pristine VO$_2$ thin film.

Figure 2(a) summarizes how oxidation time in the O$_2$-rich muffle furnace at 350°C for vanadium thin films of three different thicknesses affects the infrared transmission contrast between the insulating phase (measured at 25°C) and metallic phase (measured at 95°C). The optimal oxidation time is determined once the 54% transmittance at $\lambda$ = 8 μm line (dashed line) is reached with all the vanadium just fully oxidized, at which transmittance saturates over time. Further oxidation will lead to more over-oxidation from $VO_2$ to $V_2O_5$ at the film surface, which barely changes the infrared transmission at 25°C as both insulating $VO_2$ and $V_2O_5$ are infrared transparent but significantly increases the infrared transmission at 95°C as more $VO_2$ in lossy metallic phase is over-oxidized into transparent $V_2O_5$. Accordingly, 100 nm, 150 nm and 200 nm vanadium were found to have optimal oxidation time to be 400 mins, 950 mins, and 1700 mins. Note that, at the optimal oxidation time, the transmission contrast is not necessarily the largest due to the fact that surface over-oxides could exist in $O_2$-rich furnace atmosphere before all the vanadium within the film is oxidized. More importantly, during the oxidation process it should be ensured that all the vanadium is fully oxidized, and the transmission contrast can be maximized by surface reduction process later. As optimal oxidation time is passed, notable increase in transmittance can be observed for both insulating and metallic phase (Fig. 2b). Due to growth of transparent $V_2O_5$ with low index of $n\sim2.2$ (at $\lambda = 2$ μm)[45] on the surface, slight anti-reflection effect results in transmittance increment in shorter wavelengths ($\lambda < 8$ μm). Likewise, transmittance dips around 10 μm and 12 μm wavelengths associated with O-V-O bond and V=O bond respectively[46] enlarge over time. Note that, at the metallic phase of fully oxidized $VO_2$ thin films, there are 26%, 11%, 9% difference (at 9 μm) for 100 nm, 150 nm, and 200 nm vanadium respectively compared to that of the expected spectra. This corresponds to 24% of the expected maximized transmittance contrast ($\Delta\tau = 46\%$ for 150 nm) between insulating and metallic phase. Hence, the adverse effect of surface over-oxides cannot be neglected.

In order to achieve the best quality of thermally grown $VO_2$ thin films and thus maximize the infrared transmission contrast upon phase transition, reduction of surface over-oxides to $VO_2$ is carried out via rapid thermal processing at 500°C with forming gas (5%$H_2$ / 95%$N_2$). Reduction process with different durations at 500°C is conducted for the fully oxidized 150-nm vanadium thin film. As shown in Fig. 3(a), the infrared transmittance is measured at insulating (25°C) and metallic (95°C) phases for each reduction duration of 0 mins, 45 mins, 60 mins, 70 mins, 77.5 mins, and 80 mins. Note that 0.1 LPM forming gas is purged at atmospheric pressure throughout heating, reduction and cooling stages. Therefore, even with 0 mins reduction duration at 500°C, slight reduction could occur during heating and cooling stages at a rate of 10°C per second. This effect can be observed from 0 min reduction duration in the FTIR transmission, which decreases in the insulating phase and slightly increases in the metallic phase compared to those right after fully oxidation. The infrared transmission in the insulating phase of VO$_2$ measured at 25°C further decreases with 45 mins of reduction at 500°C, but it starts bouncing back with 60 and 70 mins of reduction, and finally reaches the expected spectrum for optimally reduced VO$_2$ film after 77.5 mins. On the other hand, the transmission in the metallic phase measured at 95°C continues to decrease slowly with longer reduction time and matches with that of optimally reduced VO$_2$ thin film after 77.5 mins of reduction. However, when the reduction time further increases by another 2.5 mins, the infrared transmittance at the insulating phase plummets from around 0.5 to 0.15. Correspondingly, Figure 3(b) shows that the transmittance contrast $\Delta \tau_\lambda$ at 9 μm wavelength between the insulating and metallic phases is about 0.29 with 0 min reduction, decreases to 0.21 after 45 mins of reduction, gradually increases to a maximum of 0.46 with optimal reduction 77.5 mins, and suddenly drops to 0.14 with 80 mins of reduction.

The evolutional behavior of the thermally oxidized vanadium thin film on reduction time can be understood by the phase diagram of different vanadium oxide states,[43] based on which it is hypothesized that high-oxides such as V$_2$O$_5$ at the film surface reduces to V$_6$O$_{13}$ first and then VO$_2$ with optimal time, and finally V$_2$O$_3$ with excessive time, as depicted in Fig. 3(b). To confirm it, Grazing Incident X-ray Diffraction scans are performed to reveal the vanadium oxide states at the film surface and within the film after different reduction durations (Fig. 4). Starting with fully oxidized (no reduction) sample (blue solid lines), as-expected high-oxides V$_2$O$_5$ peak ($2\theta = 20.4°$) is detected at both the surface and within the film. Presence of V$_3$O$_7$ ($2\theta = 55.0°$) is also observed, which overlaps with VO$_2$ (M) (220) within 0.5°. Since the existence of VO$_2$ is confirmed with FTIR measurement on insulating and metallic phase with small transmittance contrast, it is considered that the oxidation in oxygen-rich environment results high-oxides V$_2$O$_5$ and V$_3$O$_7$ at the surface of VO$_2$ film. With 500°C reduction from 0 min to 45 mins, disappearing V$_2$O$_5$ and V$_3$O$_7$ peaks and emerging V$_6$O$_{13}$ peak (50.4°) are observed in the XRD spectra on the film surface. In addition, VO$_2$ (B) phase (55.5°) shows up in the XRD spectra within the film along with multiple small VO$_2$ (M) phase peaks. Note that B-phase is a metastable VO$_2$ phase which was observed in hydrothermal reduction process as V$_2$O$_5$ → VO$_2$(B) → VO$_2$(M).[27, 47] As the phase transition temperature for both V$_6$O$_{13}$ and VO$_2$ (B) are less than 300 K,[48, 49] both behave metallic above the room temperature, which leads to decreasing transmission contrast observed in the FTIR spectra. With 45 mins of reduction, the transmittance contrast reaches its minimum as all the V$_2$O$_5$ vanishes with highest volumetric ratio of V$_6$O$_{13}$ and VO$_2$(B).

With longer reduction time, both $V_6O_{13}$ and $VO_2$(B) are further reduced to $VO_2$(M) with the transmission contrast increases as seen from the FTIR spectra with 60 and 70 mins of reduction (Fig. 3(a)). After an optimal reduction time of 77.5 mins, only $VO_2$ (M) phase is observed on the film surface and within the film at 37.1°, 42.4° and 55.6° corresponding to (200), (210), and (220) orientations without any other oxide state or phase, which is consistent with $VO_2$ fabricated via MBE, PLD and hydrothermal process.[50-52] This clearly suggests an optimal reduction of thermally grown $VO_2$ thin film with surface over-oxides to the pristine polycrystalline monoclinic $VO_2$ whose insulator-to-metal phase transition is expected to occur around 68°C.[43] As a result, transmittance contrast is maximized after 77.5 mins of optimal reduction. With further reduction, $VO_2$ starts to reduce to lower oxide $V_2O_3$ whose peak is observed in the XRD spectra after 80 mins of reduction. Note that $V_2O_3$ is metallic at room temperature with phase transition temperature around 160 K,[48] which leads to diminishing contrast in transmittance measured at 25°C and 95°C. Excessive reduction at even longer time could finally reduce the oxidized film back to vanadium completely.

In order to fully understand the evolutional IMT behaviors such as phase transition temperature and thermal hysteresis of the oxidized vanadium films reduced after different duration, comprehensive temperature-dependent optical characterizations are carried out. Figure 5 presents the infrared transmittance spectra measured between 25°C and 95°C in 5°C intervals (2°C interval within phase transition region) upon heating and cooling for the fully oxidized 150-nm vanadium thin film after (a) no reduction, (b) slight reduction with 0 min, (c) optimal reduction with 77.5 mins, and (d) over reduction with 80 mins via rapid thermal processing. Figure 6 shows the corresponding heating-cooling curves as a function of temperature for transmittance at 9 μm wavelength, as well as the derivative with respect to temperature for each case. Without reduction gradual change in transmittance spectra is observed (Fig. 5a) and phase transition takes place during heating around 66°C (where largest derivative magnitude exists) or 49°C upon cooling (Fig. 6a). The existence of over-oxides on the surface leads to not only a smaller transmittance contrast of 23% (at $\lambda$ = 9 μm) but also a wider thermal hysteresis of 17°C. Slight reduction occurs with forming gas purging during heating up and cooling down stages of rapid thermal annealing even with 0 min at 500°C, which improves the transmission contrast by 6% with similar thermal hysteresis width of 18.5°C (Fig. 5b and 6b).

With optimal reduction time of 77.5 mins, pristine $VO_2$ (M) is obtained throughout the entire film, which leads to not only the maximal transmission of 46% at $\lambda = 9$ μm but also a sharp IMT phase transition of about 15°C and a narrow thermal hysteresis of only 4°C (Fig. 6c). In particular, the derivative curve exhibits a full width at a half maximum (FWHM) of 2.2°C around 66°C upon heating and a FWHM of 2.8°C around 62°C upon cooling. The average phase transition temperature is taken as 64°C, which is slightly lower than typically reported 68°C.[43] Similar behavior has also been observed in MBE-based $VO_2$ with narrow thermal hysteresis.[16] Factors such as grain size and crystallinity could also affect hysteresis width and transition temperature slightly.[53, 54] Compared to that without any reduction, the optimal reduction process has narrowed hysteresis width by 13°C, and augmented the transmission contrast $\Delta\tau_\lambda$ at $\lambda = 9$ μm by another 23% with one-fold enhancement. For the over reduced $VO_2$ thin film after 80 mins rapid thermal annealing, due to the formation of $V_2O_3$ the infrared transmission drops (Fig. 5d) and thermal hysteresis widens to 23.5°C (Fig. 6d). The evolutional study on the reduction time clearly indicates optimal reduction is required to get rid of surface over-oxides (i.e., $V_2O_5$, $V_6O_{13}$) and avoid the formation of low-oxide ($V_2O_3$) for obtaining pristine monolithic $VO_2$ from thermal growth method with maximal transmission contrast, sharp IMT phase transition and narrow thermal hysteresis.

To confirm the evolutional IMT behaviors from fully oxidized 150-nm vanadium films reduced after different durations, temperature-dependent electrical resistivity is measured (See Sec. 4 for experimental details) between 25°C and 95°C in 5°C intervals (2°C interval within phase transition region) upon heating and cooling. Without reduction (Fig. 7a), the resistivity exhibits about 50 times of variation upon a broad IMT phase transition and a wide thermal hysteresis loop due to the existence of surface over-oxides. With 0-min rapid thermal processing (Fig. 7b), slight reduction occurs with insulating surface over-oxides start to be reduced to metallic $V_6O_{13}$ or $VO_2$, which leads to $10^2$ times of variation in resistivity upon similar broad phase transition. With the thermally grown $VO_2$ thin film after 77.5-min rapid thermal processing, the resistivity achieves a variation of $10^3$ times upon phase transition, which is comparable to that those pristine $VO_2$ fabricated by MBE or PLD techniques,[16, 55, 56] thanks to the removal of surface over-oxides from the optimal reduction. In addition, sharp phase transition and narrow thermal hysteresis are clearly seen from the resistivity derivative curves, which is consistent with the observation from the optical characterizations. Lastly, over reduction after 80-min RTP diminishes the variation of resistivity down to only 25 times upon broad phase transition due to existence of $V_2O_3$ with transition temperature at 160 K,[48] which significantly decreases the electrical resistivity from 2000 Ω·cm (pristine $VO_2$) to 30 Ω·cm at room temperature as a result of its metallic behavior.

Finally, refractive index $n$ and extinction coefficient $\kappa$ of optimally reduced 300-nm-thick polycrystalline $VO_2$ (M) thin film are extracted for both the insulating and metallic phases by fitting the measured transmittance and reflectance spectra. With the help of global optimization (See Sec. 4 for details), excellent agreement is obtained between the fitted and measured spectra, as shown in Fig. 8(a) and (b) for the insulating and metallic phase, respectively. The relative root-mean square errors (RMSE) are 1.55% (transmittance) and 2.59% (reflectance) for the insulating phase, and 11% (transmittance) and 3.82% (reflectance) for the metallic phase via the Drude model. Transmittance at metallic phase has relatively larger RMSE value due to its small values < 4%. As shown in Fig. 8(c), fitted refractive index $n$ and extinction coefficient $\kappa$ of the optimal thermally grown $VO_2$ for the insulating phase agree well with the literature data for single-crystalline $VO_2$ (ordinary component) [57] except for weaker phonon absorption in the long wavelengths, which is understandable considering its polycrystalline nature. On the other hand, the fitted ($n$, $k$) values of the optimal thermally grown $VO_2$ in the metallic phase match well with the those for single-crystalline $VO_2$ [57] as seen in Fig. 8(d), with fitted Drude parameters of high frequency constant $\varepsilon_\infty = 3.39$, plasma frequency $\omega_p = 3.16 \times 10^{15}$ rad/s, and scattering rate $\gamma = 9.17 \times 10^{14}$ rad/s. This confirms the success in removing the surface over-oxides and avoiding the formation of low-oxide with optimal reduction process via rapid thermal processing to obtain pristine $VO_2$ thin film from thermal growth method with maximal infrared transmission. See Figure S2 for comparison on infrared ($n$, $k$) values of high-quality $VO_2$ prepared by other fabrication methods such as reactive sputtering, ALD and PLD.[18, 21, 56]

3. Conclusion

In summary, this work has demonstrated the successful fabrication of pristine $VO_2$ thin film via furnace oxidation in oxygen-rich atmosphere and rapid thermal reduction of surface oxides with forming gas (5%$H_2$ / 95%$N_2$). The contrast in infrared transmission upon $VO_2$ phase transition is maximized with 2-fold enhancement compared to the oxidized vanadium film without reduction. The study on the reduction time reveals the evolutional behavior in infrared transmission variation upon phase transition which is due to hydrothermal process among different vanadium oxide states directly measured by the XRD. Temperature dependent electrical resistivity of the oxidized vanadium films reduced over different durations also shows consistent phase transition behaviors with 3 orders of magnitude drop. Sharp phase transition and narrow thermal hysteresis within 2 ~ 4°C are achieved from optimally reduced $VO_2$ thin film in both infrared transmittance and electrical resistivity, comparable to best quality of $VO_2$ prepared by other sophisticated fabrication techniques.[16, 26, 37, 50, 58] This thermal growth method could lead to scalable fabrication of pristine $VO_2$ thin films and tunable coatings for applications in adaptive thermal control,[3, 4, 23, 59-61] thermal rectification,[13, 14, 62, 63] high-resolution thermal sensing,[64] and thermal camouflage.[65-68]

4. Experimental Section

*Vanadium deposition*: Undoped double-side polished silicon wafers (280 μm thick, 2 inch in diameter, $\rho$ > 10,000 Ω·cm, (100), UniversityWafer, Inc.) are used as substrates, which are cleaned with acetone and isopropyl alcohol and dried with nitrogen gas before sample fabrication. First, vanadium (purity > 99.9% sputtering target, Kurt J. Lesker) is deposited onto the undoped silicon wafers via sputtering (Lesker PVD75 sputterer) at a rate of 0.9 Å/s with DC power of 125 W in vacuum pressure of $5 \times 10^{-6}$ Torr and process pressure of 4 mTorr with Ar gas flow. Note that a time period of 10 mins is given to reach stable deposition rate before opening the shutter for 27 mins to deposit about 150 nm vanadium thin film onto the substrate. Adequate time about 30 mins is also given for the samples to sufficiently cool down to prevent any oxidation after venting the vacuum chamber to ambient. A glass slide is utilized to cover a very small part of the silicon wafer, which creates a step after deposition for thin film thickness measurement with the profilometer (Bruker DektakXT, Fig. S3) later.

*Oxidation and reduction*: The sputtered vanadium is oxidized at 350°C in air (20.9% oxygen) for a certain amount of time in a muffle furnace (KSL-1100X, MTI Corporation). The oxidation process starts with heating at 10°C per min from room temperature to 350°C with 5 LPM nitrogen gas flow (purity 99.99%). Then dry air ($O_2$: 20-22%) is supplied at 2.5 LPM into the furnace for 10 minutes. The temperature is maintained at 350°C within ±1°C by PID control for a given amount of oxidation time, after which the sample is cooled down with heater turned off and 5 LPM nitrogen purging until it reaches the room temperature. Note that high-purity nitrogen purging is to minimize oxidation during heating and cooling stages. After furnace oxidation, surface reduction is conducted in the rapid thermal annealing system (AS-ONE 100 RTP, ANNEALSYS) at 500°C with the forming gas (5%$H_2$/ 95%$N_2$) flowing above the sample that sits on 4-inch graphite susceptor. The process starts with vacuum pumping for 10 seconds to clear out any gas residues followed by purging with forming gas at 1 LPM. Once the chamber is purged back to atmospheric pressure, the sample is heated at 10°C per second ramping rate with 0.1 LPM forming gas. After the temperature reaches 500°C, the forming gas flowrate is fixed for a given duration of reduction time. The cooling process takes place at the same forming gas flowrate with heating off until the temperature drops to 100°C.

*Material characterization*: Temperature-dependent infrared transmittance and reflectance of $VO_2$ samples are measured with an FTIR spectrometer (Nicolet iS50, Thermo Fisher Scientific) at (near) normal direction from 2 μm to 20 μm in wavelength with 4 $cm^{-1}$ resolution and each spectrum averaged over 32 scans. Aluminum mirror is used as the reference for reflection measurement in an accessory (PIKE10, PIKE Technologies). See Fig. S4 for FTIR setup. Temperature-dependent electrical resistivity is measured with a home-made four-probe station (Fig. S5) along with a source meter (Keithley 2401). The resistivity is calculated from the slope of I-V curve (*S*) and thickness (*t*) as $\rho = C\pi St/ln2$, where *C* is the geometrical factor. Sample temperature is varied with a custom built temperature stage made of a thermoelectric heater and a PID temperature controller along with K-type thermocouples. Steady state condition is achieved after the sample temperature reaches setpoint for 5 mins with fluctuation less than ±1°C. Lastly, grazing incidence X-ray diffraction (GIXRD, Rigaku SmartLab) are performed with Cu K$\alpha$ source. The 2$\theta$ scans are conducted with angle of incidence $\omega$ = 0.5° and 2.0° to probe the surface and within the thin film, respectively. The obtained diffraction patterns were compared with database (HighScore Plus) as well as literature that navigate the rapid thermal processing by forming gas.

*Extraction of complex refractive index*: The wavelength-dependent dielectric function for the insulating phase of $VO_2$ is fitted using Bayesian optimization (MATLAB, bayesopt) for refractive index *n* (4 ≥ *n* ≥ 1) and extinction coefficient $\kappa$ (10 ≥ $\kappa$ ≥ 0). The optimization takes place using the acquisition function, which integrates the positive improvement of Gaussian distribution function, acquiring next probable candidate that can update the current objective function [69] as

$$f_\lambda = \max(|\tau_{\lambda,\text{measured}} - \tau_{\lambda,\text{theory}}|, |\rho_{\lambda,\text{measured}} - \rho_{\lambda,\text{theory}}|) \tag{3}$$

where $\tau$ is the transmittance, $\rho$ is the reflectance and $\lambda$ is the wavelength. Theoretical transmittance and reflectance are calculated with transfer matrix method (TMM) incoprated with ray tracing[70] for 300-nm VO$_2$ thin film on 280-µm undoped silicon wafer. On the other hand, Drude model is fitted for vanadium and metallic phase of VO$_2$ thin films with global search (MATLAB gs) along with gradient-based minimization method (MATLAB fmincon). The optimization is conducted by providing the initial guess which is used to find parameters that minimize the objective function. With obtained population of candidates from gradient-based methods, global search utilizes scatter search algorithms to generate potential points that hunts for optimal solutions and minimal objective functions.[71] Thus, it helps to reduce locating local minimum found by gradient-based method, which requires superlative initial guess. The objective function for fitting Drude model is written as,

$$f = \sqrt{\frac{1}{N}\sum_{j=1}\left(\frac{\alpha_{\text{measured}} - \alpha_{\text{theory}}}{\alpha_{\text{measured}}}\right)^2} \tag{4}$$

where $\alpha = 1 - \tau - \rho$ is the absorptance spectrum and $N$ represents number of wavelengths. Optimization is divided into two sections. First, all three parameters of the Drude model (high frequency constant $\varepsilon_\infty$, plasma frequency $\omega_p$, scattering rate $\gamma$) and thickness $d$ are fitted together. Here, Drude model is expressed as,[70]

$$\varepsilon = \varepsilon_\infty - \frac{\omega_p^2}{\omega(\omega + i\gamma)} \tag{5}$$

Second, using the fitted plasma frequency and scattering rate, high frequency constant and thickness is adjusted by replacing absorptance $\alpha$ by transmittance $\tau$ in Eq. (4). Hence, the Drude model for both vanadium (See Fig. S6) and metallic phase of VO$_2$ (Fig. 8) is determined from transmittance and reflectance spectra measured from FTIR. The fitting is validated with 300 nm-thick SiO$_2$ using Eq. (3) to mimic VO$_2$ insulating phase and 10 nm Ag to mimic VO$_2$ metallic phase using Eq. (4) as shown in Fig. S7.

Acknowledgements

This work was supported by the National Science Foundation (NSF) under Grant No. CBET-2212342. We would like to thank ASU NanoFab and Goldwater Center for use of their nanofabrication and characterization facilities.

Conflict of Interests

The authors do not declare conflict of interest.

Author Contributions

K.A. fabricated the samples, characterized optical and electrical properties, fitted the complex refractive indices, prepared the figures, and drafted the manuscript. V.K.R. performed XRD scans. L.W. conceived the idea, supervised the work, secured the funding, and revised the manuscript. All authors have reviewed and approved the final version of the manuscript.

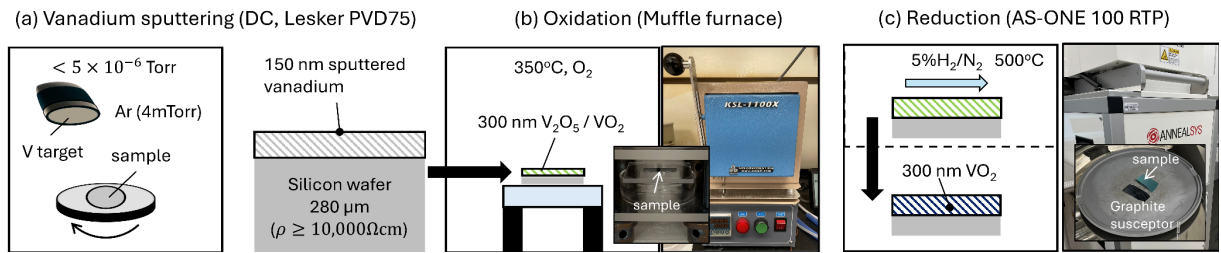

**Figure 1.** Fabrication process including (a) vanadium sputtering, (b) oxidation in $O_2$-rich muffle furnace, and (c) surface reduction in rapid thermal annealing with forming gas (5% $H_2$ / 95% $N_2$).

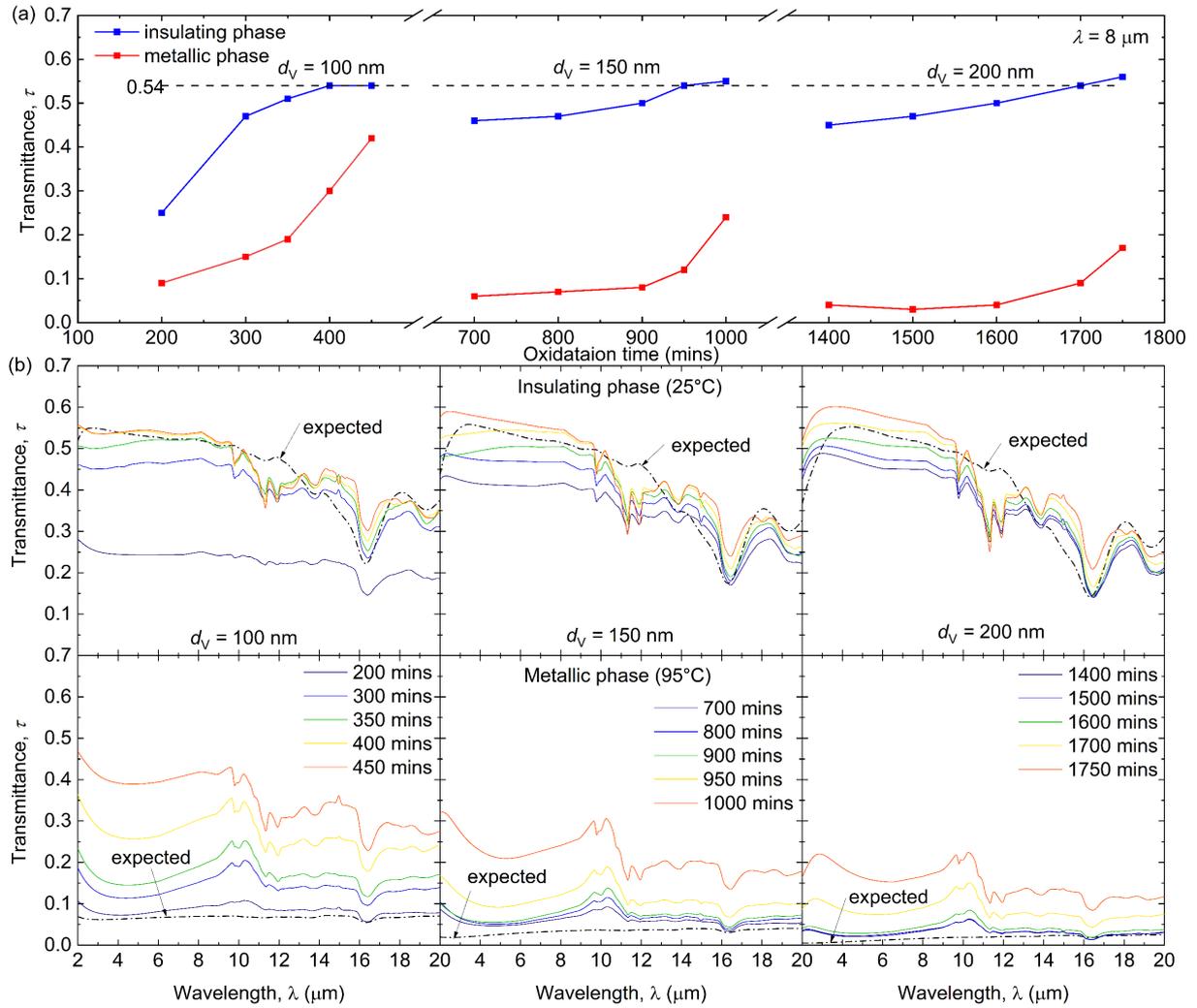

**Figure 2.** Thermal oxidation time study: (a) transmittance at $\lambda = 8$ μm with different oxidation time for vanadium thickness $d_v$ = 100 nm, 150 nm, and 200 nm; (b) transmittance spectrum at different oxidation time for both insulating (measured at 25°C) and metallic phase (measured at 95°C) for vanadium thickness $d_v$ = 100 nm, 150 nm, and 200 nm.

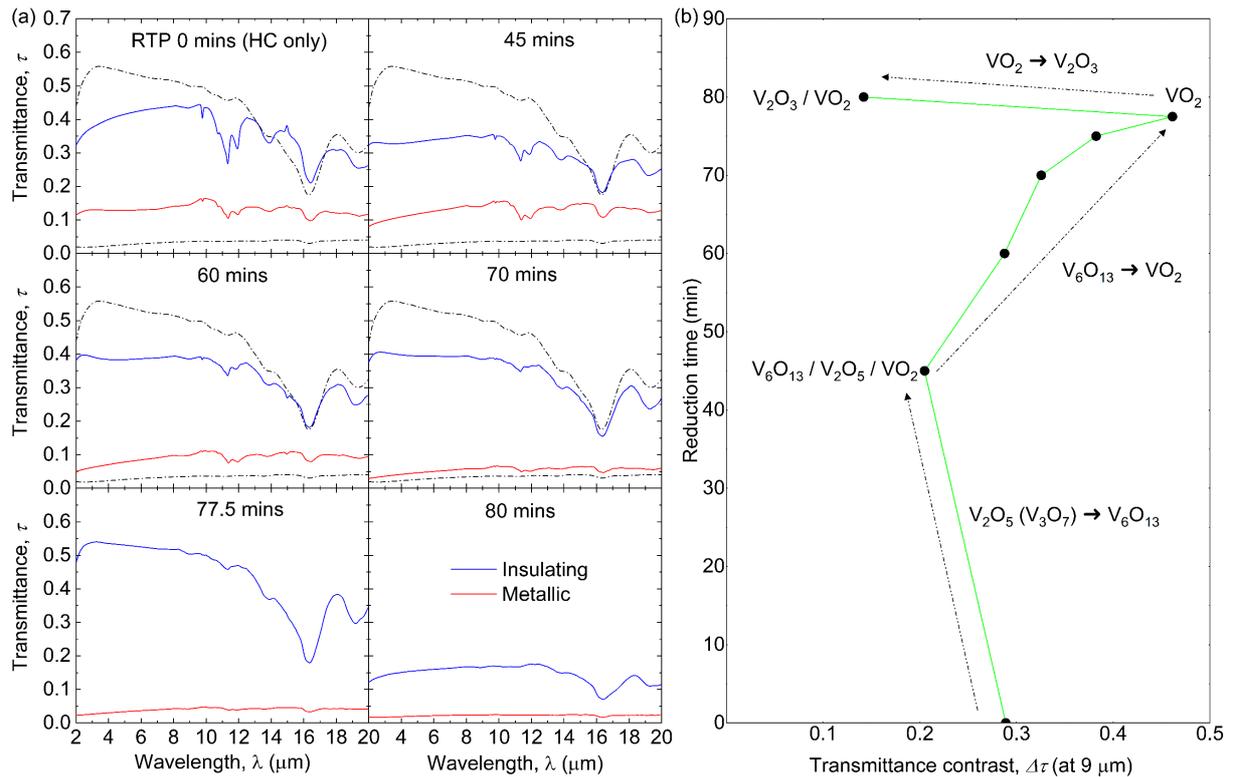

**Figure 3.** Surface reduction time study via rapid thermal annealing with forming gas for fully oxidized 150-nm vanadium thin film on undoped silicon wafer: (a) transmittance spectrum at insulating and metallic phase at different RTP reduction time. Dash lines show expected transmittance spectrum at insulating and metallic phase. (b) The relationship between transmittance contrast $\Delta\tau$ at $\lambda = 9$ μm and reduction time.

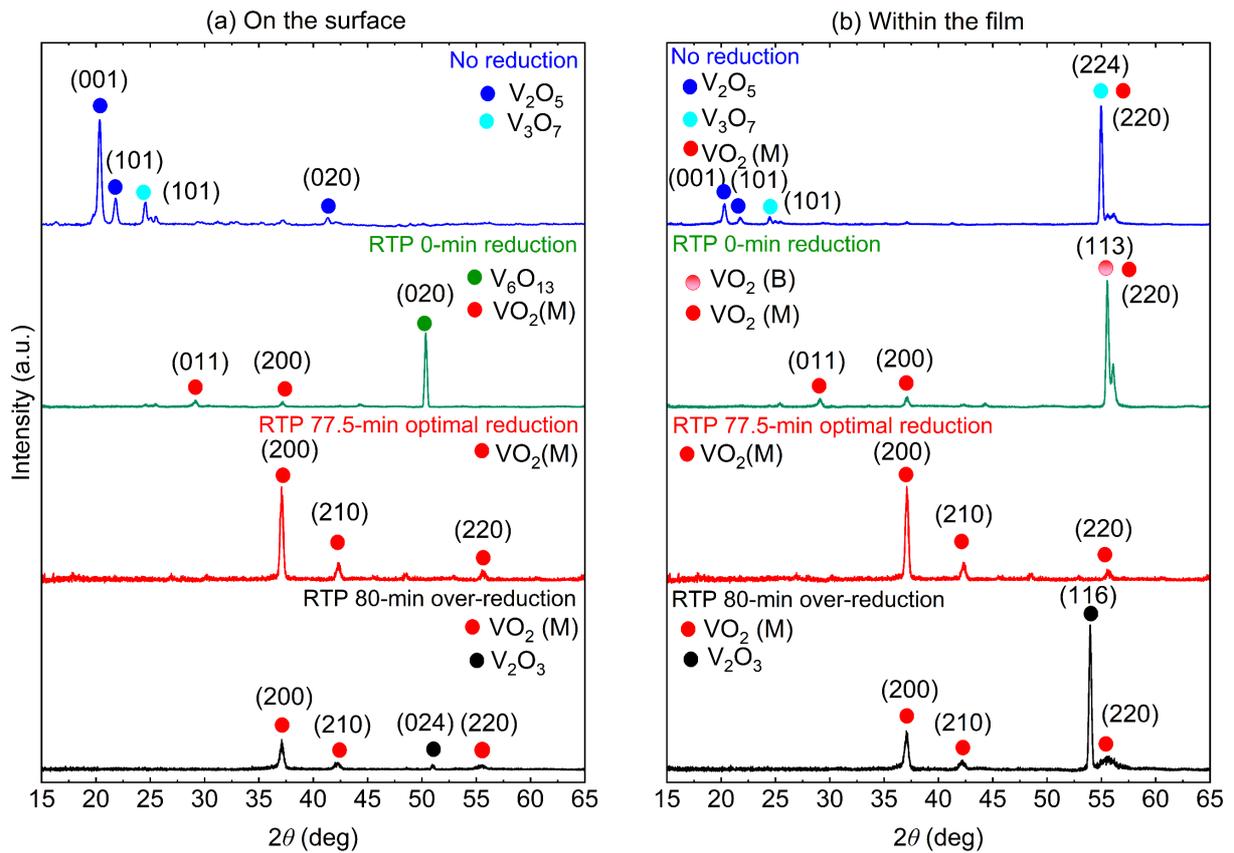

**Figure 4.** GIXRD diffraction patterns (a) on the surface (incident angle 0.5°) and (b) within the film (incident angle 2.0°) of fully oxidized 150-nm vanadium thin film on undoped silicon wafer after no reduction, slight reduction (RTP 0 min), optimal reduction (RTP 77.5 mins), over reduction (RTP 80 mins).

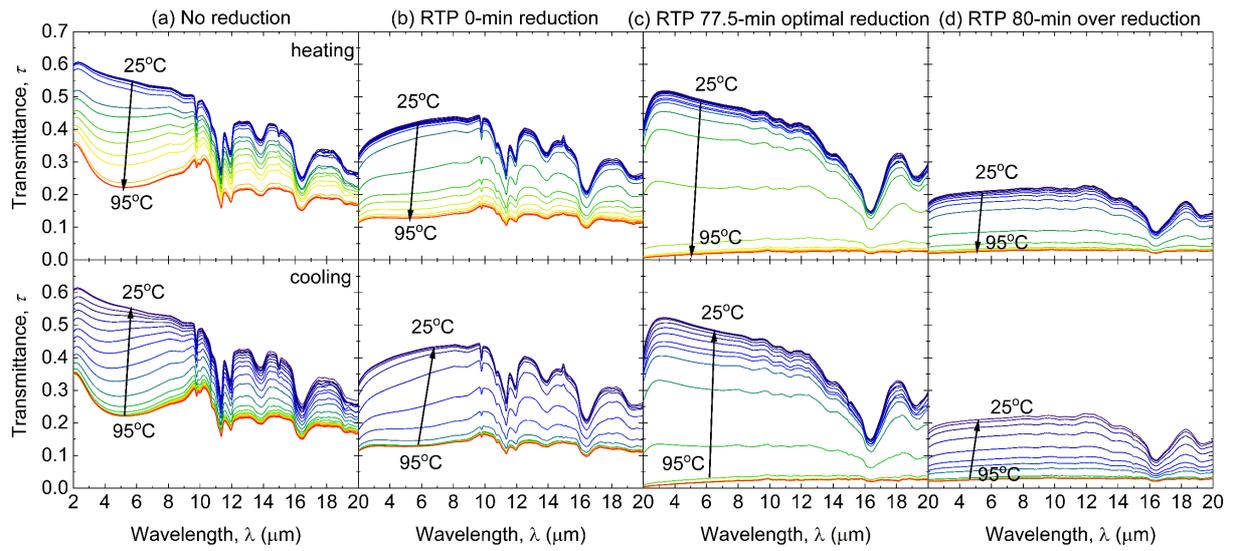

**Figure 5.** Insulator-to-metal phase transition behaviors of 150-nm vanadium thin film on undoped silicon wafer after (a) no reduction, (b) slight reduction (RTP 0 min), (c) optimal reduction (RTP 77.5 mins), (d) over reduction (RTP 80 mins): temperature-dependent transmittance spectrum during heating (top panel) and cooling (bottom panel).

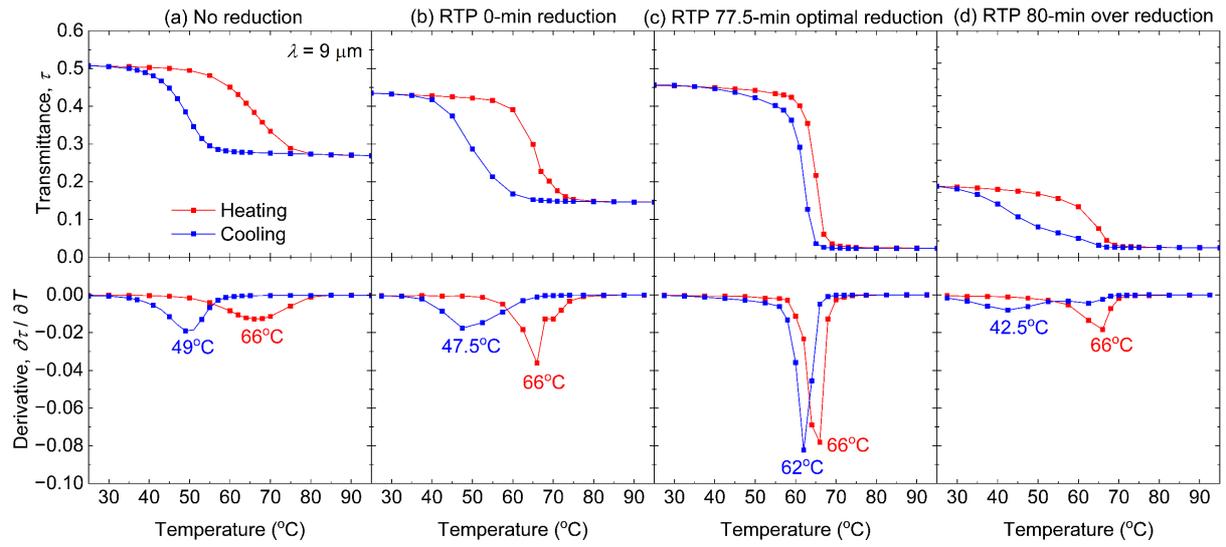

**Figure 6.** Insulator-to-metal phase transition behaviors of 150-nm vanadium thin film on undoped silicon wafer after (a) no reduction, (b) slight reduction (RTP 0 min), (c) optimal reduction (RTP 77.5 mins), (d) over reduction (RTP 80 mins): transmittance at $\lambda = 9$ μm (top panel) and its derivative (bottom panel) with respect to temperature from both heating and cooling.

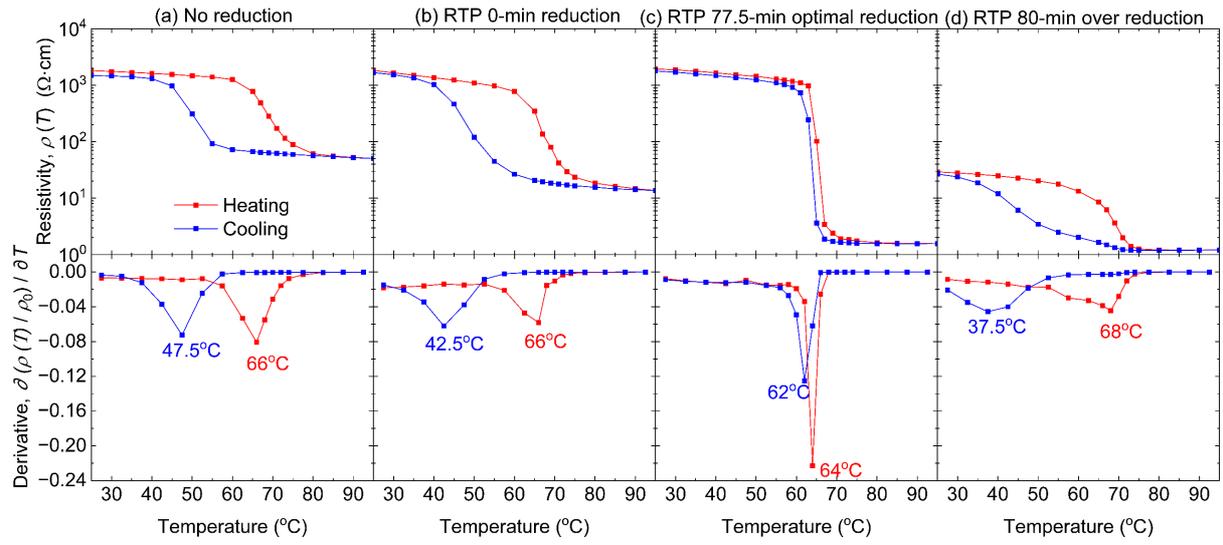

**Figure 7.** Temperature-dependent electrical resistivity (top panel) upon heating and cooling and its derivative with respect to temperature (bottom panel) of 150-nm vanadium thin film on undoped silicon wafer after (a) no reduction, (b) slight reduction (RTP 0 min), (c) optimal reduction (RTP 77.5 mins), (d) over reduction (RTP 80 mins), normalized to those measured at 25°C.

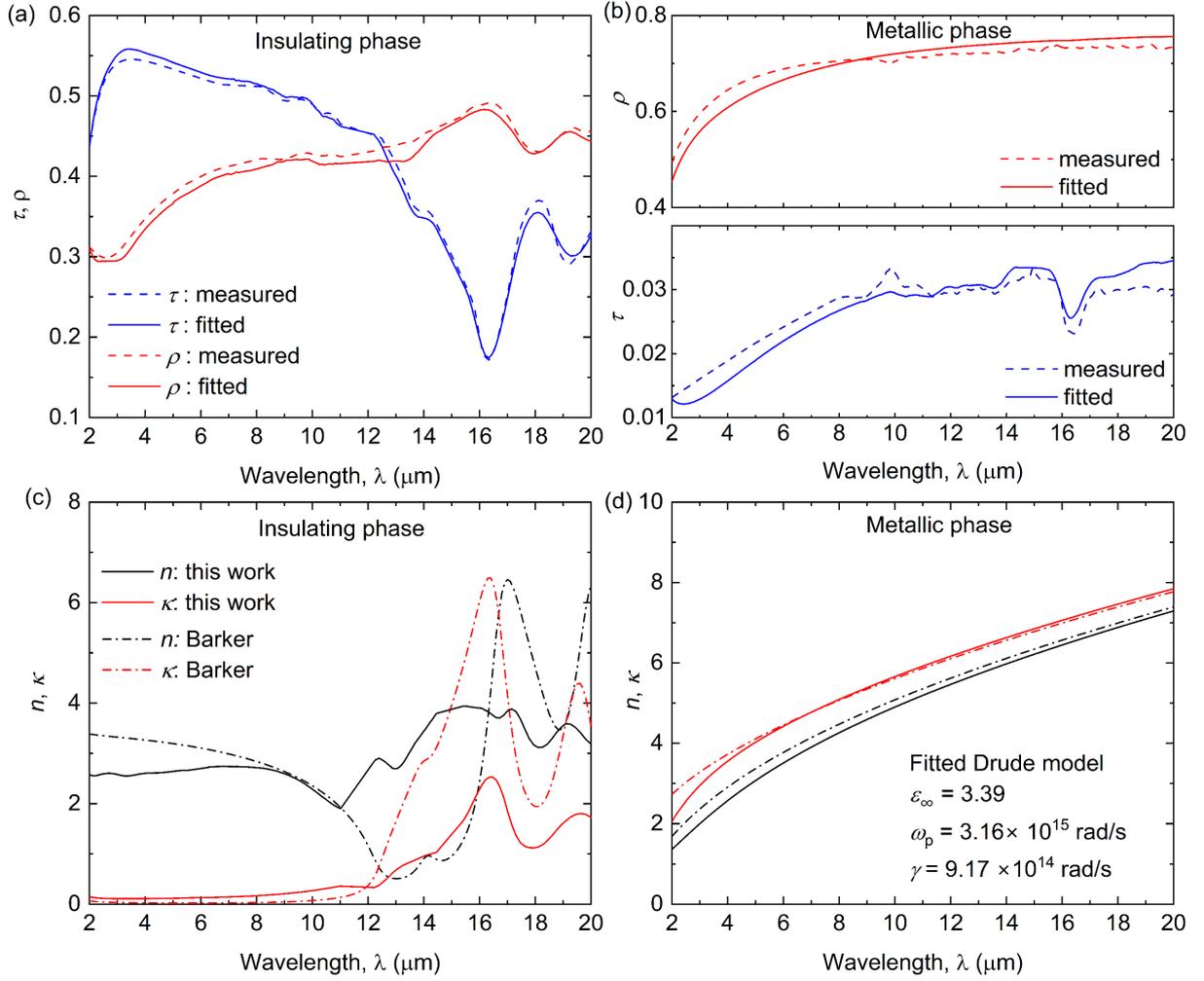

**Figure 8.** Modelled infrared transmittance ($\tau$) and reflectance ($\rho$) spectra of optimal thermally grown 300 nm VO$_2$ thin film on undoped silicon wafer in comparison with FTIR measurements at (a) insulating phase (25°C) and (b) metallic phase (95°C). Fitted refractive index (*n*) and extinction coefficient ($\kappa$) at (c) insulating phase and (d) metallic phase with the data for single crystalline VO$_2$ from Barker et al. in Ref. [56] for comparison.